\documentclass[fleqn,usenatbib]{mnras}

\usepackage[T1]{fontenc}

\DeclareRobustCommand{\VAN}[3]{#2}
\let\VANthebibliography\thebibliography
\def\thebibliography{\DeclareRobustCommand{\VAN}[3]{##3}\VANthebibliography}

\usepackage{graphicx}	
\usepackage{amsmath}	
\usepackage{amssymb}	
\usepackage{deluxetable}
\usepackage{caption,subcaption}
\graphicspath{ {./figures/} }

\usepackage{newtxtext,newtxmath}

\title[HI in Red Geysers]{The HI Content of Red Geyser Galaxies}

\author[Frank et al.]{
Emily Frank,$^{1}$\thanks{E-mail: ef202@st-andrews.ac.uk (EF)}
David V. Stark,$^{2,9}$, Karen Masters,$^{3}$,
Namrata Roy,$^{4}$, Rogério Riffel,$^{5,10}$, Ivan Lacerna,$^{6,11}$,
\newauthor Rogemar A. Riffel,$^{7,10}$, 
Dmitry Bizyaev,$^{8}$
\\
$^{1}$Department of Physics and Astronomy, Vassar College, 124 Raymond Ave, Box 743, Poughkeepsie, NY, 12604, USA\\
$^{2}$Department of Physics and Astronomy, Haverford College, 370 Lancaster Ave, Haverford, PA 19041, USA\\
$^{3}$Department of Physics and Astronomy, Haverford College, 370 Lancaster Ave, Haverford, PA 19041, USA\\
$^{4}$Department of Astronomy and Astrophysics, UC Santa Cruz, 1156 High St, Santa Cruz, CA 95064, USA\\
$^{5}$Departamento de Astronomia, Instituto de F\'\i sica,
Universidade Federal do Rio Grande do Sul, CP 15051, 91501-970, Porto
Alegre, RS, Brazil \\
$^{6}$Instituto de Astronom\'ia y Ciencias Planetarias, Universidad de Atacama, Copayapu 485, Copiap\'o, Chile\\
$^7$Departamento de F\'isica, CCNE, Universidade Federal de Santa Maria, 97105-900, Santa Maria, RS, Brazil\\
$^{8}$Department of Astronomy, New Mexico State University, 1780 E University Ave, Las Cruces, NM 88003, USA\\
$^{9}$ Department of Astronomy, University of Washington, 15th Ave. NE, Room C319, Seattle, WA 98195-0002, USA\\
$^{10}$Laborat\'orio Interinstitucional de e-Astronomia - LIneA, Rua
Gal. Jos\'e Cristino 77, Rio de Janeiro, RJ - 20921-400, Brazil\\
$^{11}$Millennium Institute of Astrophysics, Nuncio Monsenor Sotero Sanz 100, Of. 104, Providencia, Santiago, Chile \\
}

\date{Accepted 2022 December 16. Received 2022 December 16; in original form 2022 February 24}

\pubyear{2022}

\begin{document}
\label{firstpage}
\pagerange{\pageref{firstpage}--\pageref{lastpage}}
\maketitle
\begin{abstract}
  Red geysers are a specific type of quiescent galaxy, denoted by twin jets emerging from their galactic centers. These bisymmetric jets possibly inject energy and heat into the surrounding material, effectively suppressing star formation by stabilizing cool gas. In order to confirm the presence and evolutionary consequences of these jets, this paper discusses the scaling, stacking, and conversion of 21-cm HI flux data sourced from the HI-MaNGA survey into HI gas-to-stellar mass (G/S) spectra. Our samples were dominated by non-detections, or galaxies with weak HI signals, and consequently by HI upper limits. The stacking technique discussed successfully resolved emission features in both the red geyser G/S spectrum and the control sample G/S spectrum. From these stacked spectra, we find that on average, red geyser galaxies have G/S of $0.086\pm{0.011}$(random)$+0.029$(systematic), while non-red geyser galaxies of similar stellar mass have a G/S ratio of 0.039$\pm{0.018}$(random)$+0.013$(systematic). Therefore, we find no statistically significant evidence that the HI content of red geysers is different from the general quiescent population.
\end{abstract}





\section{Introduction}
\label{Introduction}

When galaxies cease star formation, they are classified as quiescent. Quiescence is typically the result of the depletion of the ingredients (i.e. cold gas) and/or physical interstellar medium (ISM) conditions (e.g., disk instability, self-shielding; \citealt{Krumholz09}, \citealt{Elmegreen11}) necessary for continued star formation. On average, galaxies with detectable cold gas are indeed star-forming, but there are notable exceptions \citep[e.g.][]{Serra2012, Gereb2016, Gereb2018, Roy2018, Parkash2019, Roy2021b}. These exceptions prompt the question: what is suppressing star formation in quiescent galaxies with substantial cold gas reservoirs? This research explores the possibility that "maintenance" or "radio" mode feedback from low/moderate luminosity active galactic nuclei (AGN) could inject energy into surrounding material, thereby preventing the formation of dense star-forming clouds \citep{Dave2020, Roy2021a}. Alternate answers to this question include ``morphological quenching" stabilizing gas disks against collapse \citep{Martig09} or recent accretion of the observed cold gas \citep{Davis2011, Lagos2015}.

Red geysers are a specific class of quiescent galaxy that are characterized by bisymmetric jets that spout from the nuclei \citep{Cheung2016}.  On average, they show significantly higher central radio flux compared to a control sample of galaxies without the characteristic optical bi-symmetric jet signatures, suggesting that red geysers are another representation or a subset of the general population of low-luminosity radio-mode AGN, and that feedback from the AGN may inhibit star formation via its interaction with the surrounding ISM \citep{Roy2018}. Only recently have detailed studies of low-luminosity radio AGN morphology and ISM interactions been performed (whereas radio-loud AGN and their role in quenching have been significantly studied). Several studies have found largely compact morphologies, athough there are some exceptions with extended jets $\ge 1\,{\rm kpc}$ in scale \citep{Baldi2018,Capetti2019, Webster2021,Baldi2021}. In some cases, radio jets have been directly associated with ionized gas or molecular outflows, providing a more direct evidence of jet/ISM interactions \citep{Jarvis2019,Venturi2021,Murthy2022}. A specific study of red geyers using LOFAR found that most exhibit compact radio morphology (although it is unclear whether or not this compactness is due to the $6\arcsec$ resolution). However, a few red geysers show resolved large-scale jets, lobes, or lopsided extended structures. These cases are typically accompanied by large velocity dispersion and/or H$\alpha$ equivalent width, although the exact nature of the interaction between the radio structure and ionized gas varies on a case-by-case basis \citep{Roy2021c}.

\cite{Cheung2016} posits that the jets inject heat and energy via winds into the surrounding cold, unstable gas in red geysers, suppressing star formation. While the interaction between their low-luminosity radio AGN and ISM is largely indirect (with a few exceptions), a number of compelling observations make red geysers worthwhile for the study of star formation suppression via AGN. The jets seen in $H\alpha$ emission that characterize red geysers align with the observed gas kinematic axis but are significantly misaligned with the stellar kinematic axis, and the ionized gas velocity fields are not well-fit by rotating disk models. The gas also shows significantly higher velocity ($\sim300\,{\rm km\,s^{-1}}$) compared to the stars ($\sim 40\,{\rm km\,s^{-1}}$) and also high gas-velocity dispersion ($\sim220\,{\rm km\,s^{-1}}$). Na D absorption in red geysers signals the presence of cool gas, but evidence of this gas is not accompanied by evidence of significant star formation \citep{Cheung2016, Roy2021b}.  Baldwin, Phillips, and Terlevich (BPT) diagrams \citep{Baldwin1981} confirm gas ionization in red geysers is primarily from post-asymptotic red giant stars and/or weak AGN, not star formation \citep{Cheung2016, Roy2018}.

\cite{Roy2021b} found an estimated mass of "cool" ( $T \sim 100 - 1000K$) gas from NaD measurements to be roughly $~ 10^8 M_{\odot}$. This implication that there might be additional cool / cold gas present detectable in other tracers motivated our work with HI observations. Demonstrating the quantity of cool gas is useful for gaining a sense of the quenching power of the 'geyser' phenomenon, that is, what amount of cold gas can be stabilized through this type of AGN feedback. This work is an investigation of the average quantity of HI gas in red geyser galaxies and the HI content of similar, quiescent (but lacking red geyser signatures) galaxies, and the results yield further information about the consequences of the unique activity of red geyser jets on star formation activity. Radio spectroscopic data from the HI-MaNGA survey \citep{Masters2019}, an HI follow-up campaign for the SDSS-IV MaNGA (Mapping Nearby Galaxies at Apache Point Observatory) survey \citep{Bundy2015}, is averaged using a stacking technique in an effort to produce an estimate of the average HI gas fraction contained in red geyser galaxies, compared to a control sample of galaxies matched in stellar mass, color, and inclination but without geyser-like jets. 

The construction and refinement of the samples are discussed in Section \ref{sec:Data and Sample Development}. Our discussion of the stacking procedure is given in \ref{Data Analysis}. Our final stacked spectrum and extracted gas fractions are given in Section~\ref{Results}. In Section \ref{Conclusions}, we summarize our findings and discuss their implications. 

\section{Sample and Data}
\label{sec:Data and Sample Development}

In order to learn whether red geyser galaxies have similar gas content as non-active quiescent galaxies, we use the \citet{Roy2021a} sample of 140 red geysers that were identified in the SDSS-IV MaNGA survey. The parameters of the red geyser classification were originally established by \cite{Cheung2016} but later improved in \citet{Roy2018}; red geyser galaxies possess red rest frame color ${\rm NUV}-r > 5$ (corresponding to typical SFRs less than $0.01\,{\rm M_{\odot}\,yr^{-1}}$), redshifts less than $z = 0.05$, discernible narrow bisymmetric features in spatially resolved EW(H$\alpha$) maps that are roughly aligned with the gas velocity field but not the stellar velocity field, and the presence of high spatially resolved gas velocities (typically a few times larger than the stellar velocities). Galaxy integrated colors, stellar masses, and redshifts are taken from the NASA Sloan Atlas\footnote{\url{http://www.nsatlas.org/}}.

\begin{figure*}
\includegraphics[width=\columnwidth]{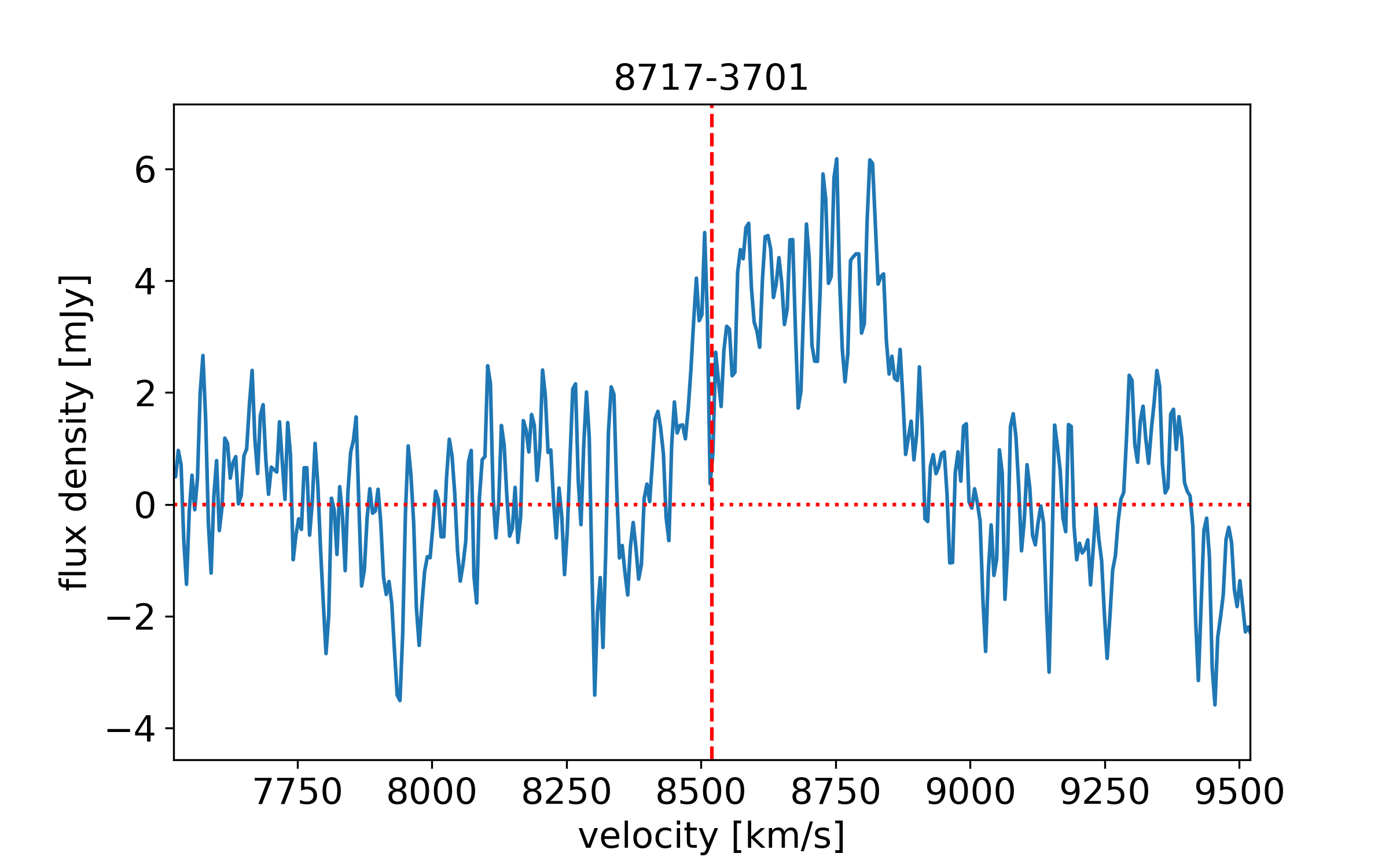}
\includegraphics[width=\columnwidth]{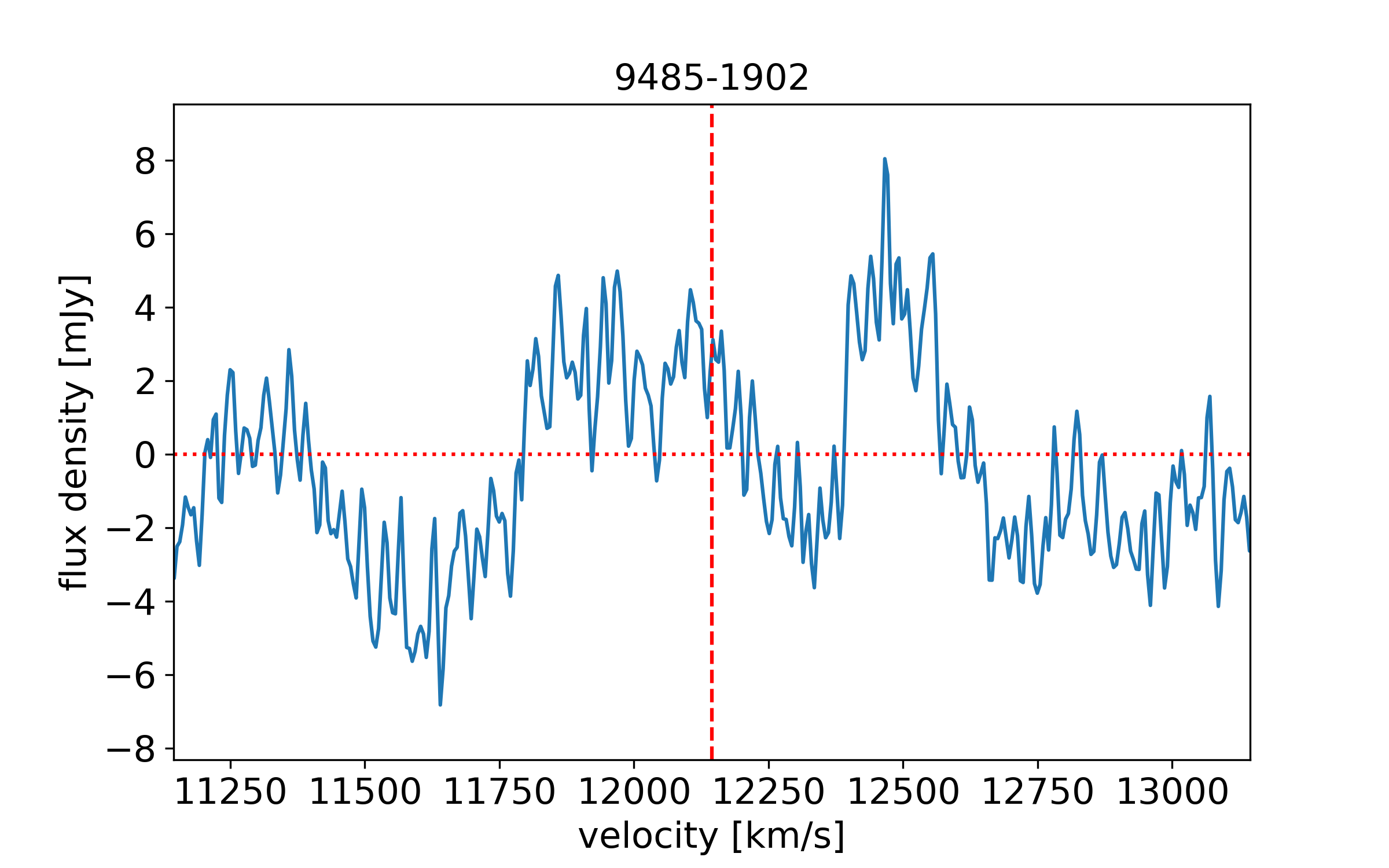}
\includegraphics[width=\columnwidth]{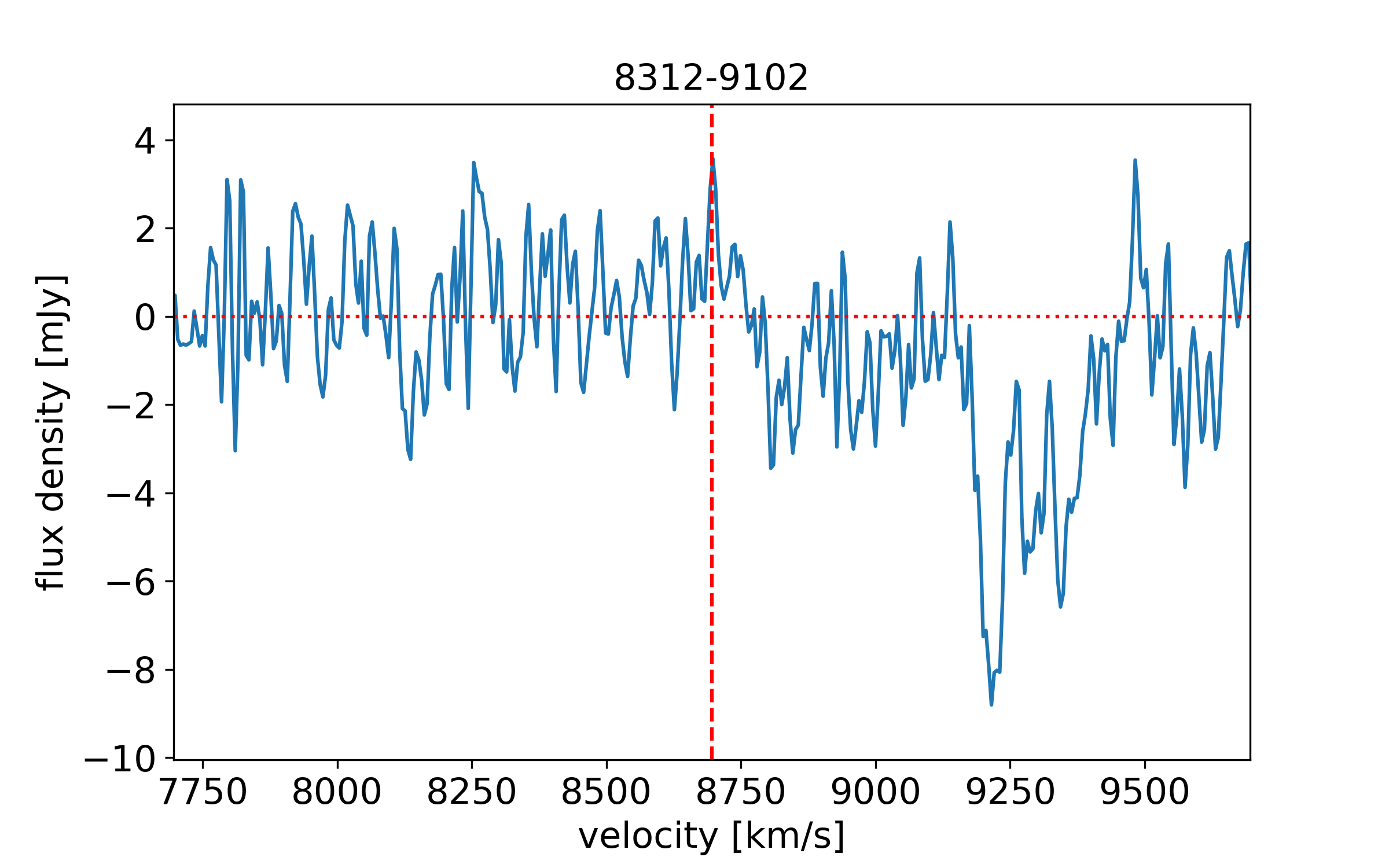}
\includegraphics[width=\columnwidth]{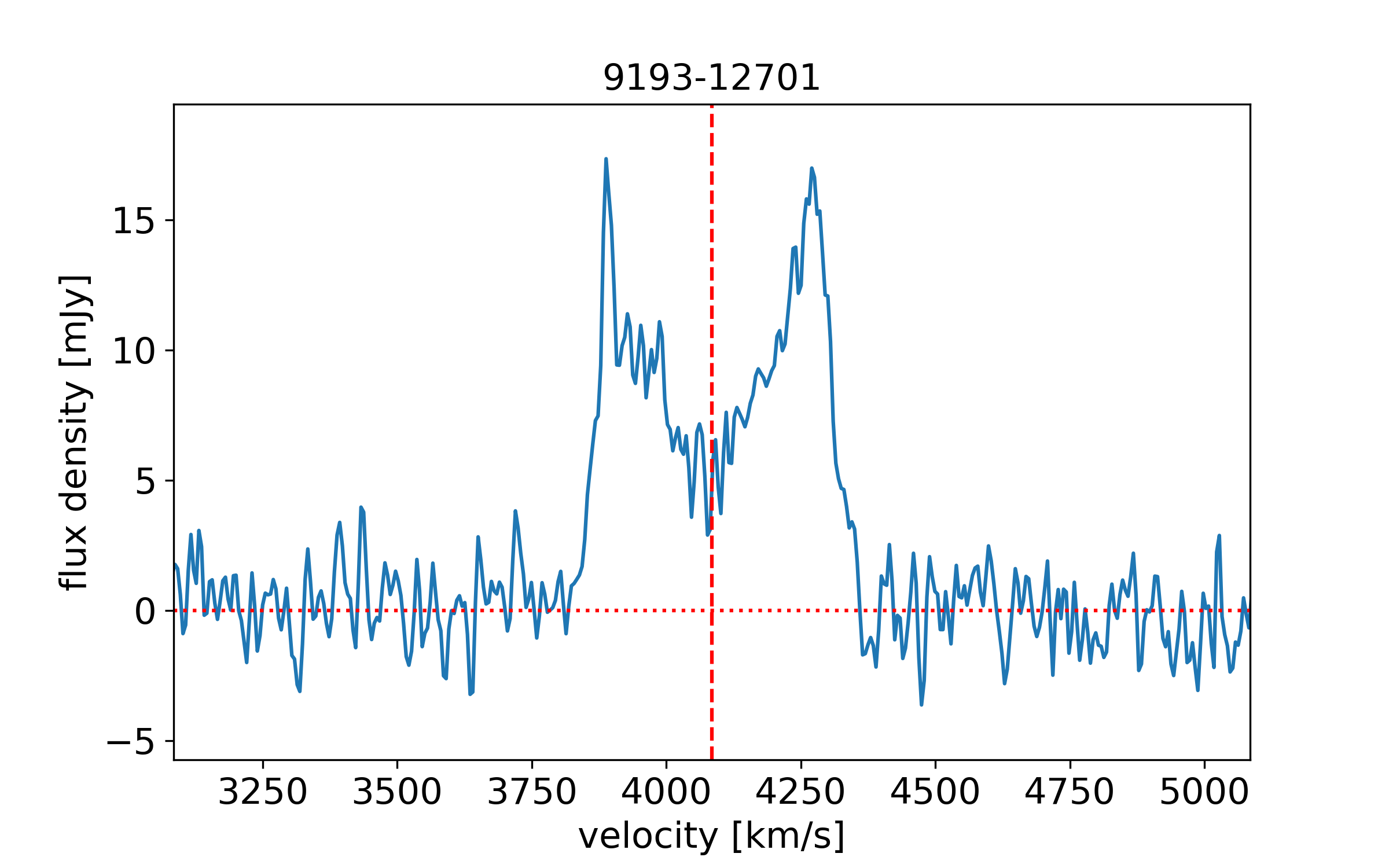}
\caption{Examples of spectra from HI-MaNGA survey. (a) There is an obvious emission in this spectrum (MANGAID 1-584637), but it is off-centered to the right. Since the spectra have been normalized to the target galaxy's redshift speed, an off-center emission implies that it is not an observation of the target galaxy.(b) MANGAID 1-121607 has features that could be HI emission-adjacent, but the baseline is not appropriately flat. (c) In this spectrum (MANGAID 1-210774) the absorption detection is the negative feature between 9150 and 9500 ${\rm km\,s^{-1}}$. Absorption detections are unique to GBT data. They are the result of an HI-detection at a similar redshift as the target during observations of blank sky. These observations (when the telescope is in the OFF position) are subtracted from the target spectra during data processing, producing a feature similar to an absorption line. (d) This galaxy (MANGAID 1-37440) displays a centered emission feature post-successful baseline fitting.
\label{fig:examplespec}}
\end{figure*}

Integrated HI spectra for this study come from the HI-MaNGA survey \citep{Masters2019, Stark2021} a 21cm follow-up program for all $z<0.05$ SDSS-IV MaNGA galaxies using the Robert C. Byrd Green Bank Telescope (GBT), and supplemented by existing 21cm observations mostly from the Arecibo Legacy Fast ALFA (ALFALFA) survey \citep{Haynes2018}. Of the 140 red geysers in our sample, there were HI spectral data for 87. At this stage, we removed 13 galaxies from our sample whose 21cm spectra exhibited unfavorable artifacts leftover from the data acquisition and reduction processes. There were a few reasons to discard these spectra: an off-center HI emission profile likely caused by a companion galaxy in the telescope beam (refer to panel a of Figure \ref{fig:examplespec}), strong baseline oscillations (refer to panel b), or a ``negative" signal (caused by 21cm signal in the reference pointing of the GBT position-switched observations) at a similar redshift as our target galaxy (refer to panel c of Fig.~\ref{fig:examplespec}). We also removed 13 galaxies with source confusion probability $>0.1$ (defined as the probability that more than 20\% of the 21cm flux in an observations comes from galaxies other than the primary target; \citealt{Stark2021}). After these sample cuts, we are left with 61 red geyser galaxies with reliable 21cm data.

\begin{figure}
\centering
\includegraphics[width=\columnwidth]{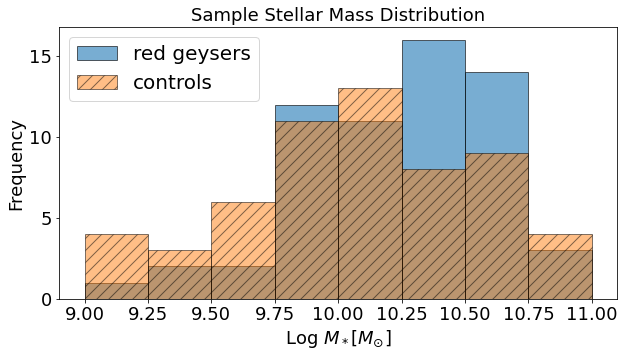}
\caption{Stellar mass distributions of primary and control samples.}
\label{figure:masshist}
\end{figure}

Probing the interplay between jets, HI content, and the suppression of star formation necessitates an analysis of the HI content of a control group of galaxies that are similarly quiescent but do not display evidence of geyser-like jets. We selected a sample of control galaxies lacking the red geyser characteristics but that adhered to the parameters for rest frame color and redshift detailed above. We further refined our selection to galaxies with stellar masses within $0.1$ dex of a sampled red geyser's stellar mass, as well as within 0.1 dex of a sampled red geyser's axial ratio. The latter parameter was included in order to avoid galaxies that might be reddened due to high inclinations, and to ensure the HI linewidths are similar. In order to maintain a stellar mass distribution comparable to the red geyser sample, the control sample consisted of 61 galaxies. The HI spectra of these galaxies were also manually inspected to remove cases with clear spectral artifacts as in Fig. \ref{fig:examplespec}. Each removed galaxy was replaced using the same matching criteria as in the construction of the original control sample. The respective mass distributions of the final samples, as shown in Figure \ref{figure:masshist}, are consistent with each other.

We do not apply weights to our samples to make them match a volume-limited sample. Therefore the mean $G/S$ measurement from our analysis of red geysers should only be considered relative to that of the control sample, not used as the true mean $G/S$ of all red geysers in the universe.


\section{Analysis Methods}
\label{Data Analysis}

The HI-MaNGA designations of HI spectra as detections or non-detections play a crucial role in analysis of the spectra. The HI-MaNGA survey handles HI flux evaluations differently depending on these designations, reporting an upper limit on HI flux ($F_{HI, lim}$)  for non-detected galaxies and an HI flux ($F_{HI,error}$) for detected galaxies. Since the HI gas-to-stellar mass ratio was derived from the analysis of stacked HI flux signals, it was essential to organize our samples manner consistent with this delineation. Both samples were largely dominated by non-detections, making up approximately 77\% of our red geyser group and 90\% of the control group.
 
\begin{figure}
\centering
\includegraphics[angle=0,scale=0.4]{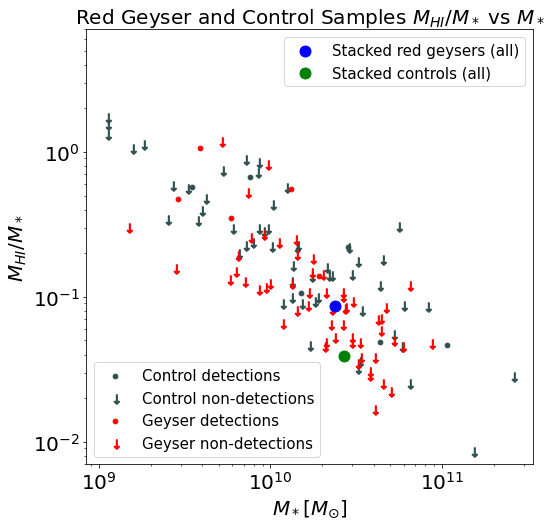}
\caption{Galaxy HI mass gas fraction against stellar mass. Individual control galaxies are depicted in grey, individual red geyser galaxies in red. The G/S of the stacked red geyser sample is depicted in blue. The G/S of the stacked control sample is depicted in green. Non-detections share down-arrow markers and dominate the plot, demonstrating the need for stacking, and detections share circle markers. The standard deviation HI mass to stellar mass is 0.304 dex and 0.319 dex for individual detections and non-detections from both samples, respectively.} 
\label{figure:mass}
\end{figure}

A preliminary evaluation of the distribution of HI gas-to-stellar mass ratios (i.e. $G/S$ ratios) as a function of stellar mass is plotted in Fig. \ref{figure:mass}. This figure illustrates how the majority of our HI data consists of non-detections (see \citealt{Stark2021} for a description of the HI upper limit calculation for individual galaxies). Since most of the detections have a large G/S ratio and low stellar mass, we explored the relationship between stellar mass and distance for the detections and non-detections in order to ensure that detections were not inherently biased towards low mass galaxies simply because they are closer. Fig \ref{figure:mass} shows that the average stellar mass of both samples are similar, and Fig \ref{figure:masshist} shows similar mass distributions between the samples. Thus any potential bias would affect both samples, and our comparative analysis would be unaffected.

It is important to note that non-detections do not preclude the presence of cold gas. \cite{Roy2018} points to $NaD$ absorption as an indicator of substantial cold gas content in red geysers, and non-detections are a consequence of both gas content and survey sensitivity (the typical HI-MaNGA HI mass sensitivity is $\sim 1 \times 10^9 {\rm M_{\odot}}$). Thus, the HI data buried in the spectra of non-detected galaxies remains consequential in the context of this project. Therefore, we estimate average $G/S$ by stacking our samples. This approach significantly increases the SNR, providing a mean signal that is more likely to be classified as a detection (and therefore produce a G/S ratio instead of an upper limit). In the case of stacking yielding a non-detection, the upper limit on the mean gas content is much stronger than for individual detections.



Our stacking procedure is as follows: First, all spectra (originally in flux density units) are scaled and converted from HI flux density to gas-to-stellar mass ratios and then weighted based on their rms noise level according to the following formula:

\begin{equation} \label{eq1}
\begin{split}
\frac{M_{HI}}{M_{*}} = 2.36 \times 10^5(d[Mpc])^2(\frac{1}{rms^2}) \frac{flux[Jy]}{M_*}
\end{split}
\end{equation}
where $d$ is distance in mega parsecs to the galaxy (found using $H_{o} = v/H$ where $v =$ galaxy velocity and $H = 70 km s^{-1} Mpc^{-1}$) and rms is the root mean square of the noise level of the spectrum as reported by the HI-MaNGA survey. This procedure creates a $G/S$ ``spectrum" where each channel represents the HI mass in that velocity channel divided by the {\it total} stellar mass $M_{*}$. When integrating any signal within this spectrum, we recover the integrated mean $G/S$ of the stack. By accounting for distance with Eq.~\ref{eq1}, the bias of HI flux and gas-to-stellar mass ratio towards nearer galaxies was eliminated. Additionally, noisier spectra are down-weighted through the division by $1/{\rm rms}^2$, ensuring an abnormally noisy spectrum does not dilute the stacked signal.

Next, each spectrum was recentered on the observed galaxy's systemic velocity (determined from optical spectroscopy) and resampled using linear interpolation onto a new velocity grid ranging from $\pm 1000\, {\rm km\,s^{-1}}$ and sampled every $5\,{\rm km\,s^{-1}}$. We did not extrapolate beyond the edge of each spectrum. Finally, each stacked G/S sample ``spectrum" was divided by the sum of the $1/{\rm rms^2}$ weights in order to obtain the final average signal. This final division is done channel by channel and accounts for some spectra being undefined in certain channels (due to e.g., overlap with the edge of the spectrum).  
 


These final stacked ``spectra" represent how the average HI flux signal of the red geyser sample and control sample translates into average HI gas-to-stellar mass ratios that can be compared.

\section{Results}
\label{Results}

\begin{figure*}
\centering
\includegraphics[width=\columnwidth]{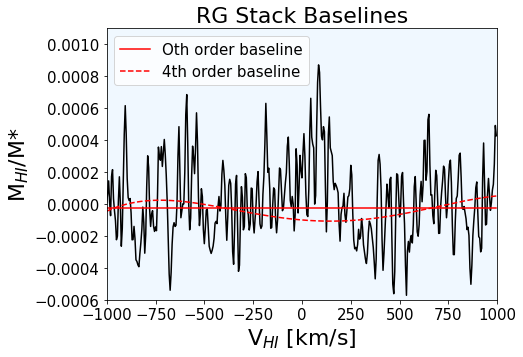}
\includegraphics[width=\columnwidth]{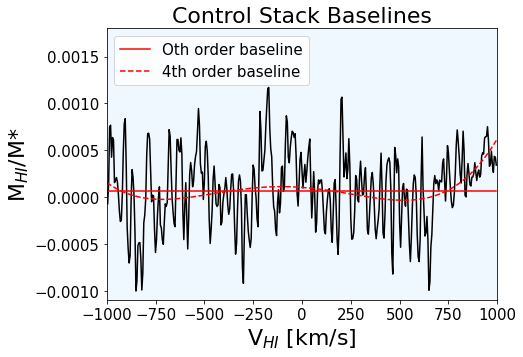}
\includegraphics[width=\columnwidth]{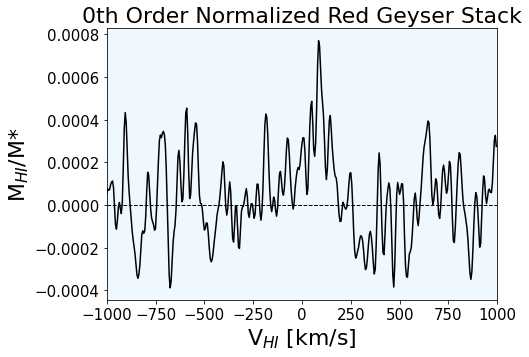}
\includegraphics[width=\columnwidth]{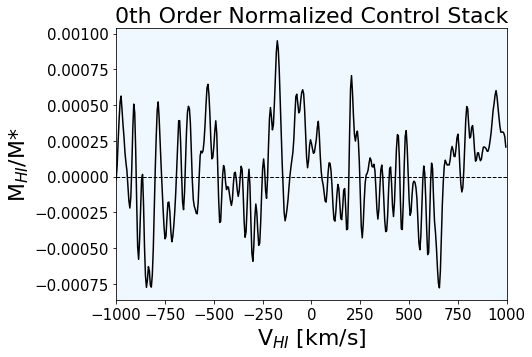}
\includegraphics[width=\columnwidth]{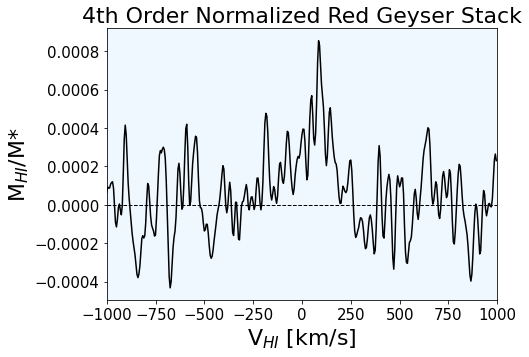}
\includegraphics[width=\columnwidth]{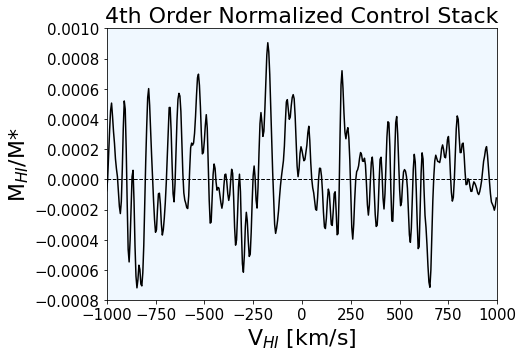}
\caption{(Top) Stacked red geysers and control spectra with zero and fourth degree baseline polynomials overlaid in solid and dashed red lines, respectively. (Middle) Final stacked spectra after zero degree baseline removal and boxcar smoothing across 4 channels (Bottom) Final stacked spectra after fourth degree baseline removal and boxcar smoothing across 4 channels.}
\label{figure:stacks}
\end{figure*}


\begin{figure}
\centering
\includegraphics[width=\columnwidth]{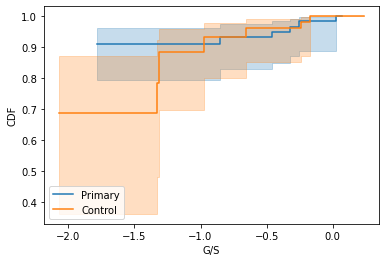}
\caption{Estimated $G/S$ cumulative distribution functions of the red geyser and control samples using the Kaplan-Meier estimator. Within uncertainties, the distributions show no clear difference. }
\label{figure:CDF}
\end{figure}




The final stacked $G/S$ ``spectra" for the red geyser and control samples are shown at the top of Fig. \ref{figure:stacks}. We visually identify weak signals in each stack, centered at approximately $V_{\rm HI}=0\,{\rm km\,s^{-1}}$. However, there is a possibility of residual baseline variations in the final stacked spectra. Before measuring properties of the stacks, we remove these baseline variations by fitting polynomials to the spectra while ignoring the central $\pm 200\,{\rm km\, s^{-1}}$ where we see the possible signal in the stacks (see Fig.~\ref{figure:stacks}). The integrated $G/S$ of the red geyser and control samples are then found by integrating the flux within $\pm 200\,{\rm km\,s^{-1}}$ of the baseline-subtracted spectrum. The final rms noise, $rms$, of each baseline-subtracted spectrum was estimated using channels with velocities outside $\pm 200\,{\rm km\,s^{-1}}$. The 1-sigma error on $G/S$ is estimated based on Equation (1) of \cite{Masters2019}:
\begin{equation} \label{eq2}
\sigma_{G/S} = rms \sqrt{\Delta v W}
\end{equation}
 where $\Delta v$ is the effective velocity resolution of the HI-MaNGA data after Hanning smoothing ($10\,{\rm km\,s^{-1}}$) and $W=400\,{\rm km\,s^{-1}}$. We characterize the strength of the detection with its integrated signal-to-noise
\begin{equation}
     (S/N)_{int} = \frac{G/S}{\sigma_{G/S}}
\end{equation}

Baseline removal can be a major source of systematic uncertainty in single-dish 21cm spectra, especially when the signals being investigated are weak, as in our case. To determine the optimal polynomial order for our baselines (i.e., one that adequately models the baseline without over-fitting), we iteratively fit polynomials between 0th and 5th order, and measure the Bayesian Information Criterion (BIC) for each, where the BIC is defined as:
\begin{equation}
BIC = k\ln{n}-2\ln{\hat{L}}
\end{equation}
where $k$ is the number of free parameters in the model, $n$ is the number of data points, and $\hat{L}$ is the maximized value of the likelihood function. Generally speaking, when comparing two models, the one with the lower BIC is preferred, although if the preferred model is more complex, its BIC should be substantially lower than the simpler model. For the Red Geyser sample, the BIC is minimized for a 0th order fit. For the Control sample, the BIC is minimized for the 4th order fit and is significantly lower than all simpler models ($\Delta{\rm BIC} < -4.4$). Adopting the 0th and 4th order baselines for the red geyser and control samples results in mean $G/S$ measurements of $0.086\pm0.0112$ and $0.039\pm0.0180$.

While we have used the BIC to choose the optimal baseline fit for each stacked spectrum, we present the results from both the 0th and 4th order fits for both samples to demonstrate the size of the systematic uncertainty due to baseline fitting. These results are provided in Table~\ref{tab:baselines}, with the bottom two rows in Figure~\ref{figure:stacks} showing the baseline fits and baseline-subtracted spectra in each case. We find that baseline uncertainties could case cause variations of $\sim30\%$ in our $G/S$ measurements. Based on our optimum baseline fits, red geysers have a mean $G/S$ that is $0.047\pm0.021$ more than the control sample (a 2.2$\sigma$ result), while adopting the results from the 0th order baseline fit for both samples implies red geysers have $0.034\pm0.023$ more gas than the control sample (a 1.5$\sigma$ result). The statistical significance of the difference in mean G/S between our two samples is marginal (2.2-sigma) at best, but systematic errors weaken this result even further. Therefore, within our uncertainties, we do not find that red geysers have any more gas than the matched control sample at high significance.

\begin{table*}
    \caption{Baseline Subtraction Results}
    \centering
    \begin{tabular}{c|c|c|c|c|c}
    \hline
     Sample & Degree of Fitted Polynomial & rms* & $G/S$ & $\sigma_{G/S}$ & $(S/N)_{\rm int}$ \\
    \hline
    Red Geyser & 4 &      1.727 $\times 10^{-4}$ &      0.115 &       1.09$\times 10^{-2}$ &       10.5 \\
    {\bf Control} &   {\bf 4} &     $\mathbf{2.84 \times 10^{-4}}$ &  {\bf 0.039} &       {\bf 1.80 $\times 10^{-2}$ }      &      {\bf 2.2}     \\
    \hline
    {\bf Red Geyser} & {\bf 0} &      $\mathbf{1.775 \times 10^{-4}}$ &     {\bf 0.086} &       $\mathbf{1.12\times 10^{-2}}$ &       {\bf 7.7} \\
    Control &   0 &    3.01 $\times 10^{-4}$ &  0.052 &       1.96 $\times 10^{-2}$       &       2.6      \\
\hline
    \end{tabular}
    \begin{itemize}
    \item []$^*$ rms of the red geyser stacked spectrum is the standard deviation around but not including the detection signal.
    \end{itemize}
    \label{tab:baselines}
\end{table*}



We also investigate whether we can infer any differences in the $G/S$ distributions of our two samples. The large fraction of upper HI limits in our samples makes plotting exact probability distributions impossible, so we instead use the Kaplan-Meier estimator \citep{Kaplan58} to estimate the survival function, $S(G/S)$, i.e., the probability that $G/S$ is greater than a given value. The cumulative distribution function is then $F(G/S)=1-S(G/S)$.

Figure \ref{figure:CDF} shows the CDFs of our red geyser and control samples with their associated 95\% confidence intervals. Within the errors, there is no clear difference in the CDFs of these two populations, although the extreme censoring fractions limit our ability to draw conclusions from this approach.

\section{Discussion and Conclusions}
\label{Conclusions}

In an effort to determine the average HI content of red geyser galaxies, a class of quiescent galaxies characterized by their lack of star formation and biconical jets seen in EW(H$\alpha$) maps, we stack 21cm spectra for 61 red geysers, finding an mean HI-to-stellar mass ratio of 0.086$\pm$0.011(random)$+$0.029(systematic). A stellar mass-matched control sample of 61 quiescent galaxies without biconical outflow signatures has a mean HI-to-stellar mass ratio of 0.039$\pm$0.018(random)$+$0.013(systematic). While we measure twice as much gas in red geysers as the control sample, this difference has only a 2.2-sigma significance using the optimal baseline fits. Systematic uncertainties from baseline fitting weaken this result further. We also find no evidence of significant differences in the distribution of G/S for these two samples using the Kaplan-Meier estimator. Therefore, we conclude there is no difference in the mean G/S for red geysers and the general quiescent population.

We stress that while we have determined the means of these two populations, the higher order properties of the distributions remain relatively unconstrained. Past work on the HI content in early-type galaxies has shown them to have an extremely broad range over several orders of magnitude \citep{Serra2012}. We caution the reader to treat our mean $G/S$ estimates as the centers of the distributions, not necessarily a highly common value or mode, for our samples. Furthermore, we have not corrected our samples to be volume-limited, so the measured $G/S$ is not necessarily representative of the true $G/S$ for red geysers in the universe.

Red geysers are hypothesized to be in a state of ``maintenance-mode feedback," where low-level AGN activity is able to deposit enough energy through jets/winds into the ISM to offset cooling and suppress star formation \citep{Cheung2016, Roy2018, Roy2021b}. We do not find definitive evidence that there is any excess or deficit of gas in red geysers compared to general quiescent galaxies, which is in agreement with broader studies of the gas content of AGN \citep[e.g., ][]{Fabello11}. This finding may lend support to the notion that red geysers reflect a intermittent phase of quiescent galaxies where energy from central supermassive black hole feedback is deposited back into the ISM, after which the AGN goes quiet. Indeed, it is generally accepted that AGN turn on and off, with active times $10^{7-9}$ yr \citep{Soltan1982,Martini2001,Yu2002,Marconi2004}. Given the low supermassive black hole accretion rates of most galaxies, we do not expect this periodic AGN activity to necessary change galaxy gas content. 

The only other study of cold gas in red geysers has been performed by \citet{Roy2021b} who finds that Na D absorption (which also reflects the presence of cool gas) are roughly twice as common in red geysers compared to a control sample, with a typical gas mass of $\sim10^{8}\,{\rm M_{\odot}}$. While their gas mass is an order of magnitude lower than ours, it is important to remember that the MaNGA optical spectroscopy data used in \cite{Roy2021b} overlaps the bright stellar disks of galaxies, while GBT data is more sensitive to the overall HI content. The discrepancy between our results is consistent with the fact that HI disks tend to be highly extended well beyond optical disks \citep{Bosma1978}. The increased rate of Na D detection in red geyers by \citet{Roy2021b} may signal that cold gas is more centrally concentrated in red geysers. Our own data hint at this possibility; the stacked profiles give the impression that the red geyser profile is narrower than the control sample profile. Given that the control sample was designed to match the axial ratio of each primary galaxy, this narrow profile is unlikely to be driven by projection effects, but rather a centralized gas distribution, which would also be consistent with an AGN being actively fed by the gas disk. This point is mostly speculative, however, as the lower $S/N$ of the control sample stack makes it difficult to reliably assess its shape.

This work represents one of the first attempts to characterize the cold gas content in red geyser galaxies. As the HI-MaNGA survey continues to add data, a larger sample of red geysers may become available, enabling a more robust assessment of red geyser gas content and possibly analysis with respect to other properties (e.g., mass). Further analysis with a larger sample or deeper observations can definitively establish whether red geysers have more or comparable amounts of gas as the general quiescent population. Additionally, examining the internal conditions of the ISM in red geysers may yield important insight into how the gas is impacted by the nuclear winds. \citet{Roy2021b} have taken a major step forward by showing strong evidence for the presence of winds interacting with the disks of two red geyser galaxies. Further analysis of the ionized gas emission spectra in MaNGA may yield additional insights into how these winds impact the mean density/temperatures of the ISM \citep{Stark2021} and suppress star formation.


\section*{Acknowledgements}

We are grateful to our referee for their careful and thoughtful review of our paper. We thank the NSF and the Keck Northeastern Astronomy Consortium (KNAC) for sponsoring this research (grants AST-1005024 and AST-1950797). We also thank Dr. Colette Salyk's for her invaluable guidance on this work. We are grateful to the ALFALFA survey team members for the high quality data they provide.

Funding for the Sloan Digital Sky 
Survey IV has been provided by the 
Alfred P. Sloan Foundation, the U.S. 
Department of Energy Office of 
Science, and the Participating 
Institutions. 

SDSS-IV acknowledges support and 
resources from the Center for High 
Performance Computing  at the 
University of Utah. The SDSS 
website is www.sdss.org.

SDSS-IV is managed by the 
Astrophysical Research Consortium 
for the Participating Institutions 
of the SDSS Collaboration including 
the Brazilian Participation Group, 
the Carnegie Institution for Science, 
Carnegie Mellon University, Center for 
Astrophysics | Harvard \& 
Smithsonian, the Chilean Participation 
Group, the French Participation Group, 
Instituto de Astrof\'isica de 
Canarias, The Johns Hopkins 
University, Kavli Institute for the 
Physics and Mathematics of the 
Universe (IPMU) / University of 
Tokyo, the Korean Participation Group, 
Lawrence Berkeley National Laboratory, 
Leibniz Institut f\"ur Astrophysik 
Potsdam (AIP),  Max-Planck-Institut 
f\"ur Astronomie (MPIA Heidelberg), 
Max-Planck-Institut f\"ur 
Astrophysik (MPA Garching), 
Max-Planck-Institut f\"ur 
Extraterrestrische Physik (MPE), 
National Astronomical Observatories of 
China, New Mexico State University, 
New York University, University of 
Notre Dame, Observat\'ario 
Nacional / MCTI, The Ohio State 
University, Pennsylvania State 
University, Shanghai 
Astronomical Observatory, United 
Kingdom Participation Group, 
Universidad Nacional Aut\'onoma 
de M\'exico, University of Arizona, 
University of Colorado Boulder, 
University of Oxford, University of 
Portsmouth, University of Utah, 
University of Virginia, University 
of Washington, University of 
Wisconsin, Vanderbilt University, 
and Yale University.

The Green Bank Observatory is a facility of the National Science Foundation operated under cooperative agreement by Associated Universities, Inc.

 The authors acknowledge the work of the
entire ALFALFA collaboration team in observing, flagging, and extracting the catalog of galaxies used in this
work.

RR thanks to Conselho Nacional de Desenvolvimento Cient\'{i}fico e
Tecnol\'ogico  ( CNPq, Proj. 311223/2020-6,  304927/2017-1 and
400352/2016-8), Funda\c{c}\~ao de amparo `a pesquisa do Rio Grande do
Sul (FAPERGS, Proj. 16/2551-0000251-7 and 19/1750-2),
Coordena\c{c}\~ao de Aperfei\c{c}oamento de Pessoal de N\'{i}vel
Superior (CAPES, Proj. 0001).

\section*{Data Availability}

HI-MaNGA data used in this study is available at \url{https://www.sdss.org/dr16/manga/hi-manga/}. Ancillary data come from the Nasa Sloan Atlas (\url{https://www.sdss.org/dr13/manga/manga-target-selection/nsa/})



\bibliographystyle{mnras}
\bibliography{bibliography} 








\bsp	
\label{lastpage}
\end{document}